\documentclass[epj]{svjour}
\usepackage{graphics}
\usepackage{amsmath}
\usepackage{hyperref}
\begin{document}
\title{Drag forces on inclusions in  classical fields with dissipative dynamics}
\author{Vincent D\'emery\inst{1} \and David S. Dean\inst{1}}

\institute{Laboratoire de Physique Th\'eorique, IRSAMC, Universit\'e de Toulouse UPS and CNRS, 31062 Toulouse Cedex 4, France}
\date{Received: date / Revised version: date}
%
\abstract{We study the drag force on uniformly moving inclusions which interact linearly with dynamical free field theories  commonly used to study soft condensed matter systems.  Drag forces are
shown to be nonlinear functions of the inclusion velocity and depend strongly on the field dynamics. 
The general results obtained can be used to explain drag forces in Ising systems and also 
predict the existence of drag forces on proteins in membranes due to couplings to various physical
parameters of the membrane such as composition, phase and height fluctuations. 
} 

\maketitle

\section{Introduction}
Quantum field theory explains how particles can interact at a distance via their coupling to a quantum field \cite{wein}. However interaction at a distance also occurs in classical systems where particles
or inclusions are coupled to classical thermal fields. For instance inclusions embedded in lipid membranes can interact due to their coupling with the membrane height fluctuations \cite{gou1993}
or with the local lipid composition \cite{sac1995}. 
As well as effective interactions between inclusions, coupling to a 
classical field can also show up in the transport properties of inclusions in these fields, notably via
drag forces which can be generated and act upon uniformly moving inclusions. The knowledge of
drag forces is important as they can be used to estimate effective transport parameters, for example
diffusion constants by using the Stokes-Einstein relation. Drag forces have been studied in a wide
variety of systems, for instance on line defects moving through liquid crystals \cite{nem} and on dislocations in layered structures  \cite{dub1983} and quasi-crystals \cite{lub1986}. As well as
being present for inclusions in a field, drag forces are also experienced by objects outside but interacting with the field, for instance magnetic force microscope tips interacting with magnetic 
substrates \cite{fus2008}. In a recent letter \cite{dem2010} we analyzed the drag on an inclusion 
which interacts with the fluctuating field, for example a magnetic field at a point which moves through
an Ising ferromagnet. Our analysis was restricted to free scalar fields undergoing a general class
of overdamped dissipative dynamics. A number of remarkable features were found for the drag 
in this class of problems (i) the average  drag force $\langle f\rangle$ is a nonlinear function of the velocity, in general it is linear for small $v$ and is characterized by a friction coefficient $\lambda
=-\lim_{v\to 0} \langle f\rangle/v$ (ii) In systems where the free field theory is critical (has a diverging correlation length) this friction coefficient can diverge and is regularized by the system size (iii) At large velocities the average force decays to zero as $\langle f\rangle\sim 1/v$. It was also found that 
numerical simulations for the drag in the Ising model could be well fitted using results for free fields (corresponding to the Gaussian approximation for the field theory of the ferromagnet). 

In this paper we will give an extended account of the results and derivations of \cite{dem2010}.  In addition we will show how the divergence of the friction coefficient $\lambda$ can be regularized
by looking at the system at a finite time after the inclusion starts to move, rather than in its steady state,
and show that it diverges as a power law in time. The fluctuations of the force about its mean value
are also analyzed and we show that the zero velocity fluctuations of the force are related to the 
linear friction coefficient via a fluctuation dissipation type relation. We also pay particular attention to  computations of drag coefficients in two dimensional systems. The reason for this is that there has been much recent interest in the diffusion constant for proteins in lipid membranes. The first theoretical computation of the diffusion constant of a protein in a lipid membrane treats the protein as a solid cylinder in a two dimensional incompressible fluid layer hydrodynamically coupled  to the surrounding bulk fluid \cite{saff1975}. 
The drag force on the fluid can be computed and one finds that, via the Stokes Einstein relation, the 
diffusion constant has a weak logarithmic dependence on the cylinder radius $a$. 
Here we explore the possibility that drag may be generated by coupling to one of the several possible physical fields  associated with the membrane, for instance  height fluctuations, thickness fluctuations, composition fluctuations, local phase fluctuations.  As mentioned above, in \cite{dem2010} it was
noted that drag forces in Ising models, which are clearly interacting models,  could be well fitted by computations based on free field theories. In general we have no explanation for this, but in this paper we have analyzed the drag forces in the one dimensional Ising model with Glauber dynamics with
a (weak) point-like magnetic field moving at constant velocity. 
We find an exact expression for $\langle f\rangle$ which turns out to have the limiting form of model A dynamics for a free Gaussian theory in the continuum limit where the correlation length is large.

\section{The free field model}

\subsection{Model definition}

In this section we will analyze the drag force exerted on an inclusion moving at constant velocity $v$
which is linearly coupled to a Gaussian or free field. We denote by $\phi({\bf r})$ a scalar field on a 
$d$ dimensional space. We write the coordinates of the system as ${\bf r} =({\bf x},z)$ where the motion of the inclusion is in the $z$ direction. The Hamiltonian for the system is given by 
\begin{equation}
H = {1\over 2}\int d{\bf r} \ \phi({\bf r})\Delta \phi({\bf r}) -hK\phi({\bf Q}(t))\label{eqH}
\end{equation}
where ${\bf Q}(t)= ({\bf 0}, vt)$ is the position of the inclusion at time $t$, $\Delta$ is 
a positive self-adjoint operator and $K$ a linear operator. The instantaneous force on the inclusion in the direction $z$ is given by
\begin{equation}
f = h{\partial\over \partial z}K\phi |_{{\bf r}={\bf Q}(t)};
\end{equation} 
i.e. it is simply the partial derivative of the total energy with respect to the movement of the inclusion 
in the direction $z$, with the field values held constant.  The energetic formulation of the way in which
the inclusion interacts with the field thus has the clear advantage, with respect to say it imposing boundary conditions, of giving an unambiguous way of computing the instantaneous force. This energetic approach was recently employed to compute the thermal Casimir force in a variety of
field theories with dissipative dynamics of the type employed here \cite{dego2009}.

Note that in  the above we have used the operator notation
\begin{equation}
\Delta v({\bf r}) = \int d{\bf r}' \Delta({\bf r}-{\bf r}') v({\bf r}').
\end{equation}

We will consider a general over-damped dissipative dynamics for the field $\phi$ which can be written in the general form
\begin{equation}
{\partial \phi({\bf r})\over \partial t} = -R{\delta H\over \delta \phi({\bf r})} + \eta({\bf r},t),
\end{equation} 
where $R$ is a positive self-adjoint dynamical operator and the noise is Gaussian, white in time, with correlation function
\begin{equation}
\langle \eta({\bf r},t) \eta({\bf r}',t')\rangle = 2T \delta(t-t')R({\bf r}-{\bf r}').
\end{equation}
This choice of correlation function obeys the fluctuation dissipation relation which ensures that
the equilibrium measure for the field is the Gibbs-Boltzmann one.

\subsection{Specific examples and applications of the free field model}
Before carrying out the general calculation we will give some examples of the sorts of field theories, interactions and dynamics that one can analyze with the formalism that follows. We start with various choices of the operator $\Delta$:
\begin{eqnarray}
\Delta({\bf r}) &=& (-\nabla^2 + m^2)\delta({\bf r}), \label{gaussian} \\
\Delta({\bf r}) &=& (\kappa\nabla^4 -\sigma\nabla^2)\delta({\bf r}).  \label{helfrich}
\end{eqnarray}
Eq. (\ref{gaussian}) corresponds to the Gaussian approximation for the Hamiltonian of a ferromagnet in the Landau theory, as such the field $\phi$ can be the local magnetization, the local composition of a binary fluid or another local order parameter. The form of Eq. (\ref{helfrich}) comes from the Helfrich Hamiltonian for a lipid bilayer
\cite{hel1973}, where $\phi$ represents the height fluctuations of the membrane about its average height. The term $\kappa$ is the bending rigidity and the term $\sigma$ is the surface tension.
Still at the static level, there are a number of choices of the coupling of the inclusion to the field
$\phi$
\begin{eqnarray}
K({\bf r}) &=& \delta({\bf r}),\label{localfield} \\
K({\bf r}) &=& {\bf d}\cdot\nabla \delta({\bf r}), \label{dipole}\\
K({\bf r}) &=& \nabla^2\delta({\bf r}).\label{kcurv}
\end{eqnarray}
The coupling (\ref{localfield}) is just a localized magnetic field, that in Eq. (\ref{dipole}) is a dipole (two fields of opposite sign close to each other) and Eq. (\ref{kcurv}) arises in lipid membrane physics and represents a coupling to the membrane curvature, tending to induce the membrane to curve upwards or inwards. This sort of coupling arises for proteins whose structures are different in the upper and lower leafs of the membrane bilayer and thus  cause the membrane to become locally curved.  For the dynamics there are a number of basic models that one can consider \cite{chai2000},
\begin{eqnarray}
R({\bf r}) &=& \delta({\bf r}) , \label{mad}\\ 
R({\bf r}) &=& -\nabla^2\delta({\bf r}) , \label{mbd}\\ 
R({\bf r}) &=& {1\over (2\pi)^2}\int d{\bf k}{\exp(i{\bf k}\cdot {\bf r})\over 4\eta |{\bf k}|}. \label{hdf} 
\end{eqnarray}
The dynamical operator of Eq. (\ref{mad}) corresponds to  the simplest form of dissipative dynamics one can write down for a system whose total order parameter ${\overline\phi}={1\over V}\int_V d{\bf r}\ \phi({\bf r})$ is not conserved (also referred to as model A dynamics). This could apply to cases such as spin systems where $\phi$ represents the local magnetization, or the 
local phase ordering parameter in lipid systems (solid, gel, liquid). The operator in Eq. (\ref{mbd}) corresponds to the simplest dynamics where the total order parameter is conserved, this would need to be the case for systems where $\phi$ represented the local chemical composition and where the total number of each type of particle is conserved (model B dynamics). The operator of Eq. (\ref{hdf}) is less general, but has a more microscopic derivation, and applies to the height fluctuations of lipid membranes driven by a surrounding fluid  of viscosity $\eta$, in Eq. (\ref{hdf}) it is defined via its (two dimensional) Fourier transform.

\subsection{General analysis of drag forces}

With the explicit choice of the 
Hamiltonian in Eq. ({\ref{eqH}) the equation of motion of the field is 
\begin{equation}
{\partial \phi({\bf r})\over \partial t} = -R\Delta\phi({\bf r}) +h RK^\dagger({\bf r}-{\bf Q}(t)) + \eta({\bf r},t)
\end{equation}
where $K^\dagger({\bf r}) = K(-{\bf r})$.

We now decompose the field into its average part and its fluctuating part
$\phi= \langle \phi\rangle +\psi$, these two components obey the evolution equations
\begin{eqnarray}
{\partial \langle \phi({\bf r})\rangle \over \partial t} &=& -R\Delta\langle \phi({\bf r})\rangle  +h RK^\dagger({\bf r}-{\bf Q}(t)) \\
{\partial \psi({\bf r})\over \partial t} &=& -R\Delta\psi({\bf r}) + \eta({\bf r},t).\label{psi}
\end{eqnarray}
We thus see that the mean value of the field $\phi$ depends on the position of the inclusion but the 
fluctuations about this mean value are independent of the inclusion.  We now write the inclusion 
position as ${\bf Q}(t) = ({\bf 0},vt)$ and we write the mean value of the field $\phi$ as
\begin{equation}
\langle \phi({\bf r}, t)\rangle = g({\bf x}, z-vt,t).
\end{equation}
In the coordinate system ${\bf r} = ({\bf x}, z'=z-vt)$ the equation for $g$ is
\begin{equation}
{\partial g\over \partial t}-v{\partial g\over \partial z'}= -R\Delta g + hRK^\dagger({\bf r}) .\label{avev}
\end{equation}
The Fourier transform of $g$ defined as 
 \begin{equation}
 \tilde g({\bf k}) = \int d{\bf r} g({\bf r}) \exp(-i{\bf k}\cdot{\bf r})
 \end{equation}
obeys \cite{comm}
\begin{equation}
{\partial \tilde g\over \partial t}-ik_z v \tilde g= -\tilde R \tilde \Delta g + h\tilde R\tilde K^\dagger.
\end{equation}
In the steady state regime we can ignore the temporal derivative above and find
\begin{equation}
\tilde g({\bf k}) = {h\tilde R({\bf k}) \tilde K(-{\bf k})\over \tilde R({\bf k}) \tilde \Delta({\bf k})-ik_zv}.\label{ftgs}
\end{equation}
In the coordinate system moving with the inclusion the force is given by
\begin{equation}
\langle f\rangle = h {\partial \over \partial z'}Kg |_{{\bf r}=0}={h\over (2\pi)^d}\int d{\bf k} ik_z\tilde K({\bf k})\tilde g({\bf k}),\label{fft}
\end{equation}
and putting Eqs. ({\ref{ftgs}) and (\ref{fft}) together then yields
\begin{equation}
\langle f\rangle =  {h^2\over (2\pi)^d}\int d{\bf k} {ik_z \tilde R({\bf k})\tilde K({\bf k}) \tilde K(-{\bf k})\over \tilde R({\bf k}) \tilde \Delta({\bf k})-ik_zv}.\label{stat}
\end{equation}
For small $v$ this gives
\begin{equation}
\langle f\rangle = -\lambda v
\end{equation}
where the coefficient of friction $\lambda$ is given by
\begin{equation}\label{dragcoeff}
\lambda =  {h^2\over (2\pi)^d}\int d{\bf k} {k_z^2  \tilde K({\bf k}) \tilde K(-{\bf k})\over \tilde R({\bf k})
\tilde \Delta({\bf k})^2}.
\end{equation}
In the case where the system is isotropic, that is $\tilde K$, $\tilde R$ and $\tilde \Delta$ are functions
of $k=|{\bf k}|$ we find
\begin{equation}\label{dragcoefiso}
\lambda =  {h^2\over (2\pi)^d d }\int d{\bf k} {k^2  \tilde K(k)^2\over \tilde R(k)
\tilde \Delta(k)^2}.
\end{equation}

We can also analyze the case where the insertion is inserted at a time $t=0$ and see how the force 
evolves in time. This case is especially interesting when the corresponding steady state quantities turn
out to be divergent. Here it is convenient to work with the Laplace transform of the average force 
defined as $\overline f(s) =\int_0^\infty dt\ \exp(-st) f(t)$. Using the fact that $f(0)=0$ we can solve 
Eq. (\ref{avev}) by Laplace transforming to give
\begin{equation}
\langle \overline f(s) \rangle =  {h^2\over s(2\pi)^d}\int d{\bf k} {ik_z \tilde R({\bf k})\tilde K({\bf k}) \tilde K(-{\bf k})\over \tilde R({\bf k}) \tilde \Delta({\bf k})+s-ik_zv}.\label{flt}
\end{equation}
The static result Eq. (\ref{stat}), when it is finite,  is recovered from the pole at $s=0$ in Eq. (\ref{flt}). In the limit of small $v$ we can define a time dependent friction coefficient $\lambda(t)$ via $\langle f(t)
\rangle =-\lambda(t) v$. The Laplace transform of $\lambda(t)$ is then given by
\begin{equation}
\overline \lambda(s) =  {h^2\over s (2\pi)^d}\int d{\bf k} {k_z^2 \tilde R({\bf k})  \tilde K({\bf k}) \tilde K(-{\bf k})\over (\tilde R({\bf k})
\tilde \Delta({\bf k})+s)^2}
\end{equation}
and when the system is isotropic this can be written as

\begin{equation}\label{dragcoefflap}
\overline \lambda(s) =  \frac{h^2}{s (2\pi)^d d}\int d{\bf k} \frac{k^2 \tilde R({ k})  \tilde K^2(k)}{(\tilde R( k)\tilde \Delta({k})+s)^2}.
\end{equation}

Finally it is interesting to ask under what conditions a force can be generated in a direction
perpendicular to the direction of the insertion's uniform motion in two or more dimensions. The calculations above  can be easily extended to show that the force in the direction $x$ say is given by
\begin{equation}
\langle f_{\perp}\rangle =  {h^2\over (2\pi)^d}\int d{\bf k} {ik_x \tilde R({\bf k})\tilde K({\bf k}) \tilde K(-{\bf k})\over \tilde R({\bf k}) \tilde \Delta({\bf k})-ik_zv}.\label{dyn}
\end{equation} 
Note that this Hall like effect can be analyzed for the forces on vortices in superconductors, however
the evaluation of the force in this magnetic context requires a subtle analysis of the time 
dependent Ginzburg-Landau equations \cite{vor}. In our problem the interaction between the 
inclusions and the field is via a pure potential so the evaluation of the corresponding forces is
much more straight forward.  In an isotropic system it is clear that $\langle f_{\perp}\rangle =0$. The perpendicular friction coefficent
is given via $\langle f_{\perp}\rangle\sim -\lambda_{\perp}v$ for small $v$ as
\begin{equation}
\lambda_{\perp}= {h^2\over (2\pi)^d}\int d{\bf k} {k_x k_z \tilde K({\bf k}) \tilde K(-{\bf k})\over \tilde R({\bf k}) \tilde \Delta^2({\bf k})}.
\end{equation}
An interesting example of where this perpendicular friction coefficient can be non zero is where the
interaction term takes a dipolar form $\tilde K({\bf k}) =-i{\bf d}\cdot {\bf k}$, while the other operators remain isotropic, in this case we find
\begin{equation}
\lambda_{\perp}= {2h^2d_x d_z\over (2\pi)^d}\int d{\bf k} {k^2_x k^2_z \over \tilde R(k) \tilde \Delta^2(k)},
\end{equation}
and thus see that it can be non zero when the dipole has non zero  components in the direction of the
motion and perpendicular to the motion. For a fixed dipole modulus, 
the magnitude of the perpendicular force is maximal when the dipole is orientated at $45^o$ to the direction of the movement.  We will demonstrate the existence of this rather odd perpendicular force
later on in simulations of the two dimensional Ising model.

\subsection{Regularization of divergences in the model}

The integrals appearing in  Eqs. (\ref{stat}) and (\ref{dragcoeff}) may diverge. To be more specific, we will focus on the friction coefficient for an isotropic system. The divergences depend on  the dimension $d$ of the system and the operators $\Delta$, $R$ and $K$. For small $k$  we will take them of the form
\begin{eqnarray}
 \tilde \Delta(k) & \sim &\ k^\delta\ \label{formDelta}\\
 \tilde R(k) & \sim &\ k^\rho  \\
 \tilde K(k) & \sim  &\ k^\alpha\ . \label{formK}
\end{eqnarray}
In this notation, when the field theory has a finite correlation length $\xi =1/m$ we thus
have  $\delta=0$. 
We find that the integral in  Eq. (\ref{dragcoeff}) is infra red divergent when $d<d_c$ with $d_c$ given
by  
\begin{equation}\label{critdim}
d_c=2\delta+\rho-2\alpha-2.
\end{equation}
We  note that $d_c$ increases (i) as $\delta$ increases, i.e. long distance excitations cost less
energy, (ii) $\rho$ decreases, i.e. long distance modes relax more quickly (iii) when $\alpha$
decreases, i.e. when the coupling of the inclusion to the field is long range. 
In the case where the drag coefficient is infra red divergent it is regularized by
cutting off the $k$ integration at an infra red cut off $k_{\rm min} = \pi/L$ where $L$ is the linear 
system size.  For $d<d_c$ we find 
\begin{equation}
\lambda \sim L^{d_c-d}.
\end{equation}

As should be expected the divergence of the friction coefficient also shows up in a non-analytic
behavior of the average drag force at small $v$ and one can show that
\begin{equation}
\langle f\rangle \sim v^{1-{d_c-d\over \rho+\delta-1}}
\end{equation}
when $d<d_c$, under the conditions $\rho +\delta > 1$ and $(d_c-d)/(\rho+\delta -1) < 2$.

Finally there is another way to regularize the infrared divergence; we can measure the friction coefficient at a finite time. We expect that the friction coefficient will grow with the time as 
\begin{equation}
\lambda(t)\sim t^\phi
\end{equation}
and one can compute the exponent $\phi$ using the Laplace transform (\ref{dragcoefflap}). Making the change  of variable  $k=s^{1/(\rho+\delta)}q$, and noting that the Laplace transform of $t^\phi$ is proportional to $s^{-(1+\phi)}$, we obtain
\begin{equation}\label{divexp}
\phi=\frac{d_c-d}{\rho+\delta}\ ,
\end{equation}
where again we assume that $\rho+\delta > 1$.

In general the expressions given above for the drag force can also exhibit ultra violet divergence
which must be regularized. There are two possible physical length scales which regularize the
corresponding integrals

(i) the field theory has a natural  cut off $k= \pi/a_0$ where $a_0$ is a length scale below which the field does not fluctuate or beyond which its fluctuations are strongly suppressed. This cut off scale can be 
imposed by hand and taken to correspond to a molecular scale, for example the lipid size in lipid membrane bilayers, or because the Hamiltonian function $\tilde \Delta$ has corrections at higher order
in $k$ than its low $k$ form given in Eq. (\ref{formK}). 

(ii) the size of the inclusion $a$ gives a
cut off $k=\pi/a$, for instance instead of having a point like magnetic field inclusion where
$\tilde K = h$, one can have a Gaussian distributed field smeared over a region of size 
$a$ with $\tilde K = h\exp(-{k^2a^2\over 2})$. This means that the $k$ integration is effectively cut off
at $k=\pi/a$. For the purposes of this paper therefore we will take the cut off to be $k_{\rm max} =
{\rm min}\{{\pi/ a_0},\  {\pi/ a}\}$. However in most cases of interest is is usually $a$ which is
the larger of these two ultra violet length scales. 

The conclusion of this analysis is that when $d<d_c$ the results we obtain are dominated by the long distance properties of the theory and we see a diverging friction coefficient as $\xi\to 0$. However if
$d>d_c$ the friction coefficient becomes strongly dependent on the ultra violet cut off, for instance on
the size of the inclusion. This ultra violet dominated regime thus lacks the universality of the
infra red dominated regime and we must be careful in our choice of model and regularization to obtain physically meaningful results.

\subsection{Force fluctuations}

Here we will consider the statistical properties of the fluctuations of the force about its mean value.
Depending on the system these fluctuations may be measurable and could provide a method for
determining some of the effective parameters describing the system. We define the fluctuating component of the force as
\begin{equation}
\Delta f = f-\langle f\rangle.
\end{equation}
This fluctuating component can be written in terms of the fluctuating component of the field
$\psi$ defined in Eq. (\ref{psi}) and is given by
\begin{equation}
\Delta f =  h{\partial\over \partial z}K\psi |_{{\bf r}={\bf Q}(t)}.
\end{equation} 
In the steady state regime the correlation function of the field $\psi$ is given by 
\begin{equation}
\begin{array}{l}
\langle \psi({\bf r},t) \psi({\bf r}',t')\rangle = C({\bf r}-{\bf r}',t,t') \\
\hspace{10mm}=\displaystyle{T\int d{\bf u} \ \Delta({\bf r}-{\bf u}) \exp(-|t-t'|\Delta R)({\bf u}-{\bf r}')}.
\end{array}
\end{equation}
Using this we find that the correlation function for the force fluctuation is given by
\begin{equation}
\begin{array}{l}
\langle \Delta f(t) \Delta f(t')\rangle =
\displaystyle{{Th^2\over (2\pi)^d}\int d{\bf k} \ k_z^2 {{\tilde K}({\bf k}){\tilde K}(-{\bf k})\over {\tilde \Delta}({\bf k})}}
\\ \hspace{1cm}\times \exp\left(-|t-t'|\tilde \Delta({\bf k})\tilde R({\bf k})+ ik_z v(t-t')\right).
\end{array}
\end{equation}
The equal time correlation function is thus given by
\begin{equation}
\langle \Delta f(0) \Delta f(0)\rangle =
{Th^2\over (2\pi)^d}\int d{\bf k} \ k_z^2 {{\tilde K}({\bf k}){\tilde K}(-{\bf k})\over {\tilde \Delta}({\bf k})}.
\end{equation}
and is independent of the velocity $v$. We also see that there is a fluctuation dissipation like
relation relating the zero velocity force fluctuations to the linear friction coefficient:
\begin{equation}
\int_0^\infty dt\langle   \Delta f(t)\Delta f(0)\rangle|_{v=0}
= T\lambda\ .\label{fdt}
\end{equation}
In general any measurement of a force will not be instantaneous and will depend on the temporal
resolution of the experimental set up and will thus represent  force averaged 
over a characteristic time scale $T_m$ associated with the force measurement apparatus. 
We define the temporally averaged force over the time
window $T_m$ as
\begin{equation}
f_m = {1\over T_m}\int_0^{T_m} f(t) dt
\end{equation}
clearly we have $\langle f_m\rangle = \langle f\rangle$ and the variance at zero 
velocity $\sigma_m^2
=\langle (f_m-\langle f\rangle)^2\rangle|_{v=0}$
is given, for large $T_m$, by
\begin{equation}
\sigma_m^2 = {2T\lambda\over T_m}.
\end{equation}

\section{Numerical simulations of drag in the Ising model}

In this section we perform numerical simulations of drag forces on inclusions in the Ising model. This is an example of an interacting theory where the drag forces predicted in free field theories should also
occur. Our simulations in fact show that, despite their approximative nature in the context of 
interacting theories, our results for the free Gaussian ferromagnet account well for the phenomenology of drag observed in Ising systems. We  will consider the model on a $d$-dimensional cubic lattice of spacing $a_0$, with periodic boundary conditions, and denote by $N$ the total number of sites and spins. The Hamiltonian is given by
\begin{equation}
H=-J\sum_{(i,j)}S_iS_j-h\sum_i K_{i-i_0}S_i
\end{equation}
where $J>0$ is a ferromagnetic coupling between nearest neighbour spins and where 
$hK_{i-i_0}$ is the local field at site $i$ due to the inclusion whose position is denoted by  $i_0$.
Here the vector $K_i$ is the discrete version of the operator $K$ of Section 2. In what follows, in order to fully investigate the various models discussed in the paper, but to keep to a reasonable length and 
minimize the amount of computation time, we will restrict our study to one and two dimensions.

The system dynamics is defined in the following manner: $N$ elementary evolutions are performed during one unit of time. An elementary evolution consists of:
\begin{itemize}
\item choosing a spin set (the way of choosing it depends on the dynamics and will be given below),
\item computing the energy change $\Delta H$ associated with flipping this set of spins,
\item flipping the spins  with probability \\ $p_f=1/(1+\exp(\Delta H/T))$ or leaving them unchanged with probability $1-p_f$.
\end{itemize}
We simulate two types of dynamics:
\begin{itemize}
\item Non-conserved dynamics: only one spin is chosen randomly at each step; thus the total magnetization is not conserved. This choice is referred to as Glauber dynamics.
\item Conserved dynamics: at each step, two spins of opposite sign are randomly chosen; the total magnetization is conserved. This is a form of Kawasaki dynamics.
\end{itemize}

\par The inclusion moves in the $z$ direction with velocity $v$, so $i_{0,z}(t)={\rm int}(vt/a_0)$
(int denoting the integer part) i.e. it performs one step every $a_0/v$ units of time. To measure the force in the $z$ direction in a given configuration of the spins system, we compute the energy $H_+$ if the inclusion was at the position $i_0+{\bf z}$  and $H_-$ if it was at the position $i_0-{\bf z}$, where ${\bf z}$ is the lattice link vector in the  direction $z$. The instantaneous force is then $f=(H_--H_+)/2a_0$. Explicitly 
\begin{equation}
f(t)=\frac{1}{2a_0}\sum_i[K_{i-(i_0+{\bf z})}-K_{i-(i_0-{\bf z})}]S_i.
\end{equation}
As we are interested in the average force, we average over all Monte-Carlo (MC) time steps after 
first achieving a steady state in the simulation:
\begin{equation}
\langle f\rangle  =\lim_{T_{MC}\rightarrow\infty}\frac{1}{T_{MC}}\sum_{t=1}^{T_{MC}}f(t)\ .
\end{equation}
\par We will present two kinds of plots of the results of our simulations (i) the average magnetization in the rest frame  of the inclusion to see how the inclusion polarizes the spins around it and how this polarization cloud is deformed by the inclusion's movement (ii)  the average drag force as a function of the velocity (i.e. $\langle f \rangle (v)$). Note that the polarization induced by the inclusion is basically a generalization of the polaron in solid state physics; the polaron is the response of a body's polarization field due to the presence of an electron and modifies the dynamical properties of the electron \cite{polaron}.

\subsection{Point like magnetic fields in one and two dimensions}

Here we study the case where the inclusion creates a point like magnetic field: $K_{i-i_0}=\delta_{i,i_0}$. We take $h<0$, so that the average magnetization is (positively) proportional to the potential seen by the inclusion. 

Figure \ref{profil1d} shows the  magnetization profile for Glauber and Kawasaki dynamics, for four different speeds (with parameters $\beta=1$, $J=1$ and $h=-10$). When the particle is at rest, it sees a spherically symmetric potential, thus is does not experience any net force. As the velocity increases, the profile becomes asymmetric and its amplitude decreases: the system has  less  time to react to the presence of the inclusion and thus the polaron is deformed and becomes weaker. The main differences between the two dynamics are (i) the magnetization far from the particle is not zero with Kawasaki dynamics, because the total magnetization must remain zero; (ii) the polaron deformation appears to be larger with Kawasaki dynamics. The mean force is plotted against the velocity in Fig. \ref{dragv1d}, this figure shows that the force has a linear dependence on $v$ for small $v$, reaches a maximum and then decreases as $1/v$ for large $v$. This behavior is in agreement with our general results for free fields: the asymmetric profile is responsible for the force and for small $v$ the asymmetry increases with $v$ whereas for large $v$ the profile amplitude diminishes. As the deformation is larger for Kawasaki dynamics, the force is larger. The fits with our analytical results for model A and B dynamics are performed by varying three parameters: the cut-off and a dilatation for each axis.\par The 2d simulations give similar results (Figs. \ref{profil2d}, \ref{dragv2d}). In two dimensions we are in the high temperature regime before the ferromagnetic phase transition  ($\beta=1$, $J=0.4$, $h=-6.66$).

\begin{figure*}
\begin{center}
\resizebox{1\hsize}{!}{\includegraphics{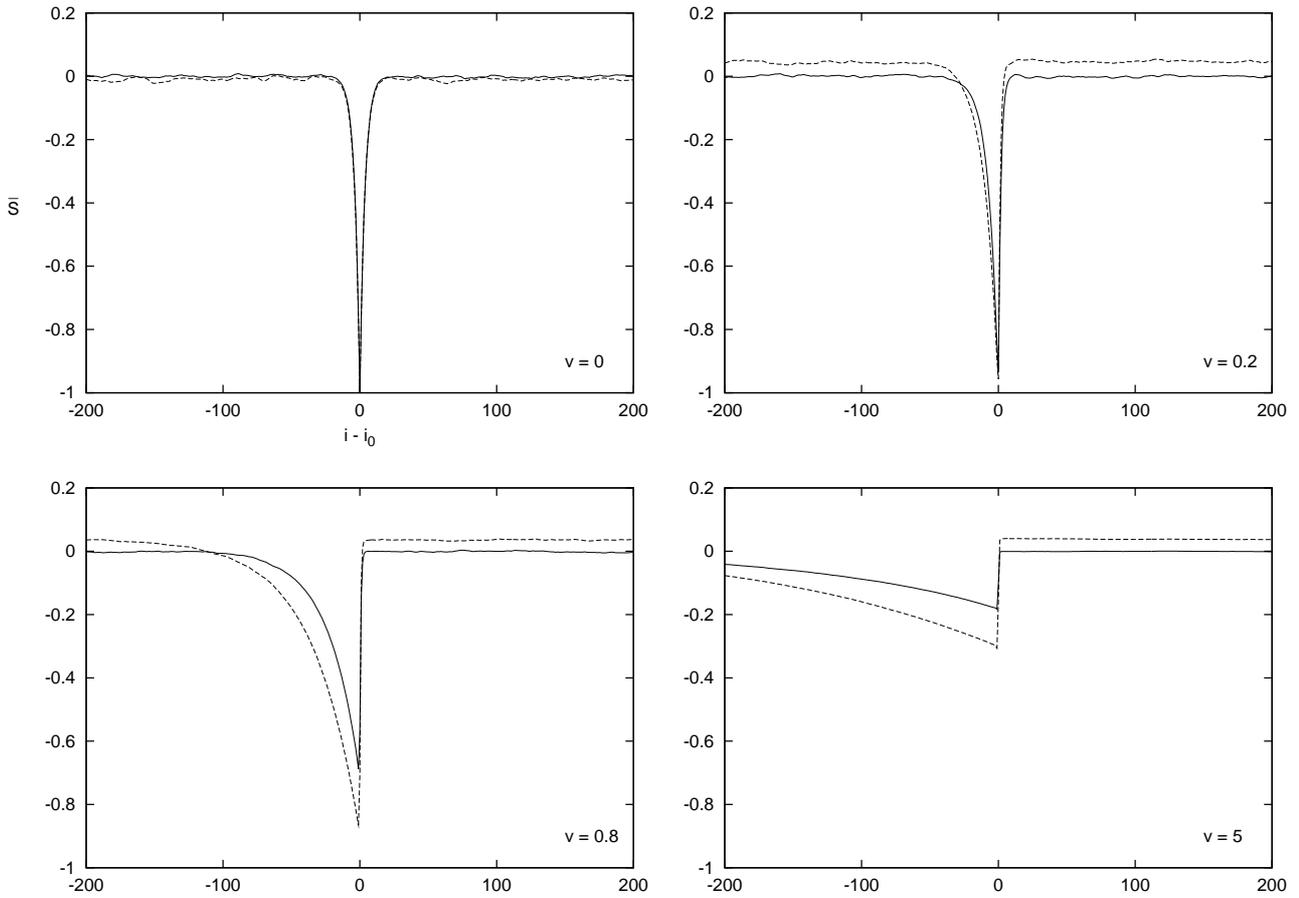}}
\end{center}
\caption{Magnetization profile for the 1d Ising model about a local magnetic
field at a single point moving with velocity $v$, for Glauber (solid lines) and Kawasaki (dashed lines) dynamics.}
\label{profil1d}
\end{figure*}

\begin{figure*}
\begin{center}
\resizebox{1\hsize}{!}{\includegraphics{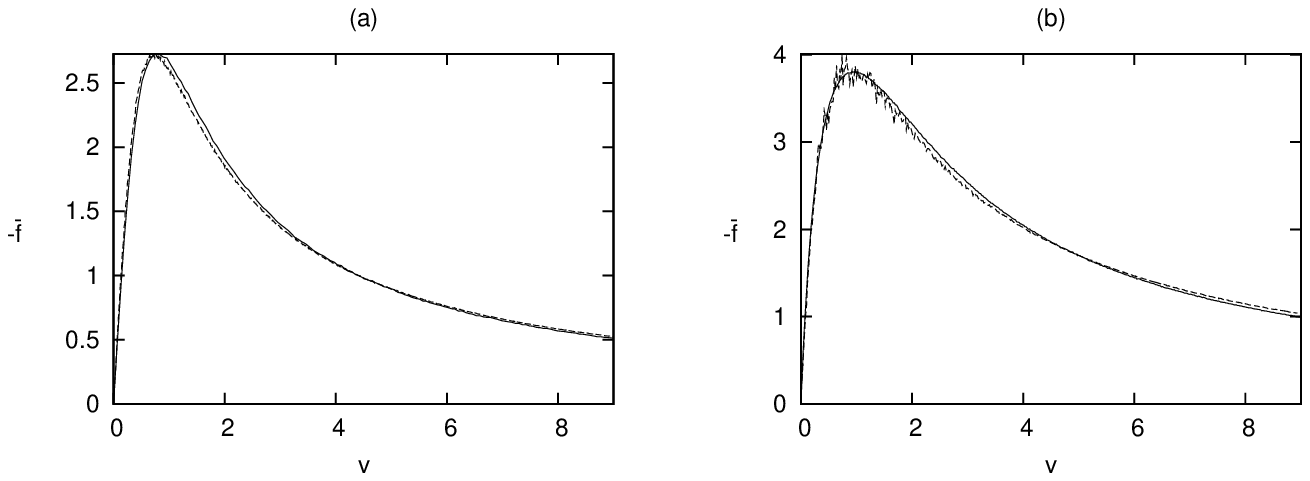}}
\end{center}
\caption{Dashed lines: average drag force $\overline f$ in the 1d Ising model  as a function of $v$ for Glauber  (a) and Kawasaki dynamics (b). Solid lines are the fits of model A (a) and model B (b) dynamics
for the Gaussian ferromagnet.}
\label{dragv1d}
\end{figure*}

 \begin{figure*}
\begin{center}
\resizebox{1\hsize}{!}{\includegraphics{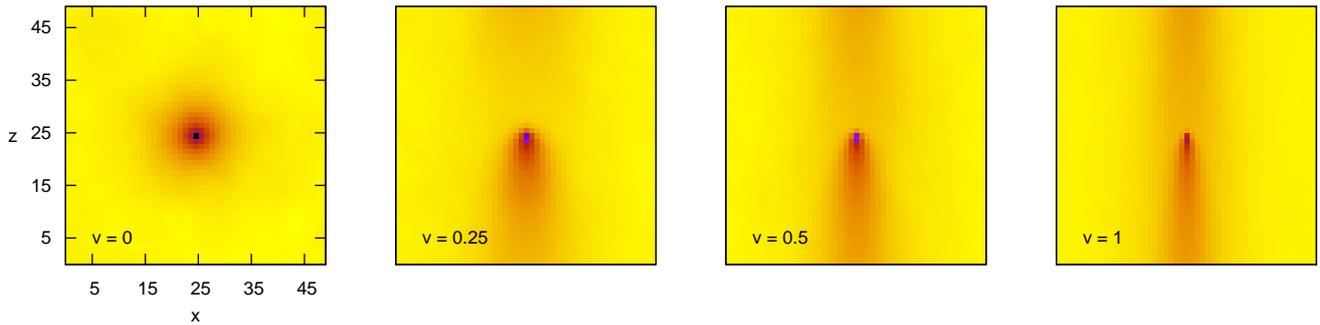}}
\end{center}
\caption{Contour plot (color online) of the magnetization profile (polaron) for the 2d Ising model with Glauber dynamics about a local magnetic
field at a single point moving with velocity $v$. 
 (high temperature phase: $\beta=1$, $J=0.4$, $h=-6.66$). }
\label{profil2d}
\end{figure*}

\begin{figure*}
\begin{center}
\resizebox{1\hsize}{!}{\includegraphics{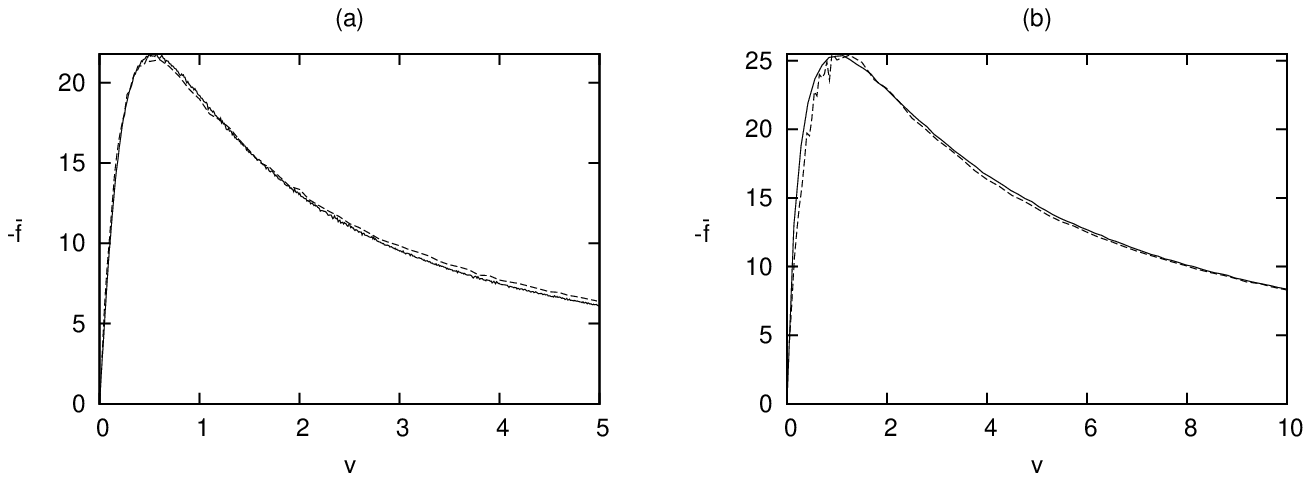}}
\end{center}
\caption{Dashed lines: average drag force $\overline f$ in the 2d Ising model  as a function of $v$ for Glauber  (a) and Kawasaki dynamics (b). Solid lines are the fits of model A (a) and model B (b) dynamics
for the Gaussian ferromagnet ($\beta=1$, $J=0.4$, $h=-6.66$). }
\label{dragv2d}
\end{figure*}

\subsection{Dipoles in two dimensions}

Here, the inclusion interacts  as a dipole: $K_{i-i_0}=(\delta_{i,i_0}-\delta_{i,i_0-{\bf u}})$, where ${\bf u}$ is a unit vector giving the direction of the dipole. Fig. \ref{dipparperp} shows the drag force for dipoles perpendicular (${\bf u}={\bf x}$) and parallel (${\bf u}={\bf z}$) to the direction of motion for Glauber dynamics. We also show the  contour plots for the local magnetization profile for both cases, at
velocities $v=0$ and $v=0.5$. As seen from the force curves, at slow speed the force does not depend on the orientation, when the speed increases the force for the perpendicular dipole becomes larger. \par Finally, we compare the forces parallel and perpendicular to the motion for a dipole orientated at $45^o$ to the direction of the motion (${\bf u}={\bf x}+{\bf z}$) in Fig. \ref{dip45}. The force in the direction {\bf x} is calculated in the same way as that described above for the force in the direction {\bf z}. We see that the transverse force has the same order of magnitude as the longitudinal force, and the same general form.
Also shown on the right is the corresponding contour plot of the local magnetization generated by the dipole at rest and for $v =0.5$. Whereas at $v=0$ the magnetization profile appears antisymmetric
about the direction of the dipole, when the dipole moves it experiences a force which pushes it to the left
on the bottom right figure of Fig. \ref{dipparperp}. This is because the leading component of the dipole barely sees the polarization created by the lower component, whereas the magnetization created by the leading component pushes away the lower component.


 \begin{figure*}
 \begin{center}
\resizebox{1\hsize}{!}{\includegraphics{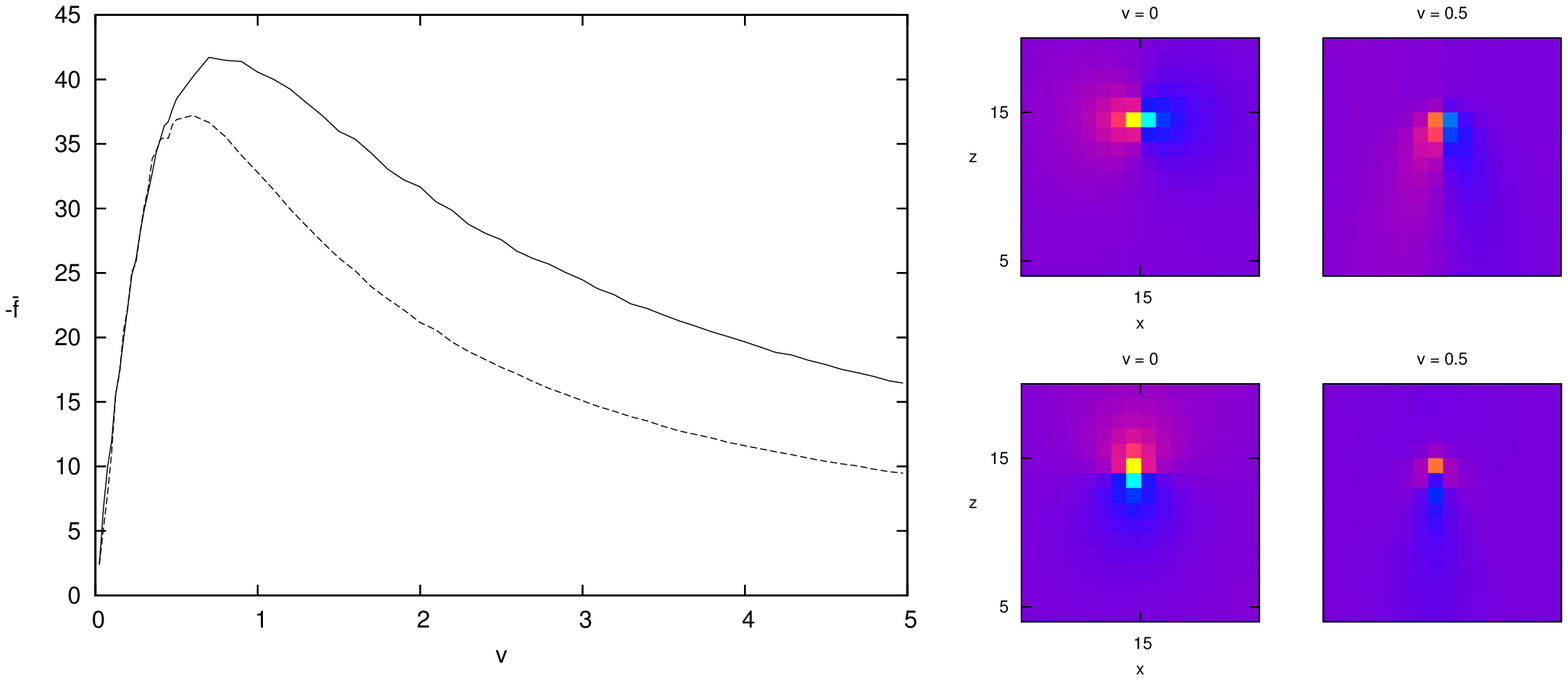}}
\end{center}
\caption{Drag force for dipoles perpendicular (solid line) and parallel (dashed line) to the motion for Glauber dynamics ($\beta=1$, $J=0.4$, $h=-6.66$); contour plot (color online) for two values of $v$ for dipoles perpendicular (first line) and parallel (second line) to the motion.}
\label{dipparperp}
\end{figure*}

 \begin{figure*}
 \begin{center}
\resizebox{1\hsize}{!}{\includegraphics{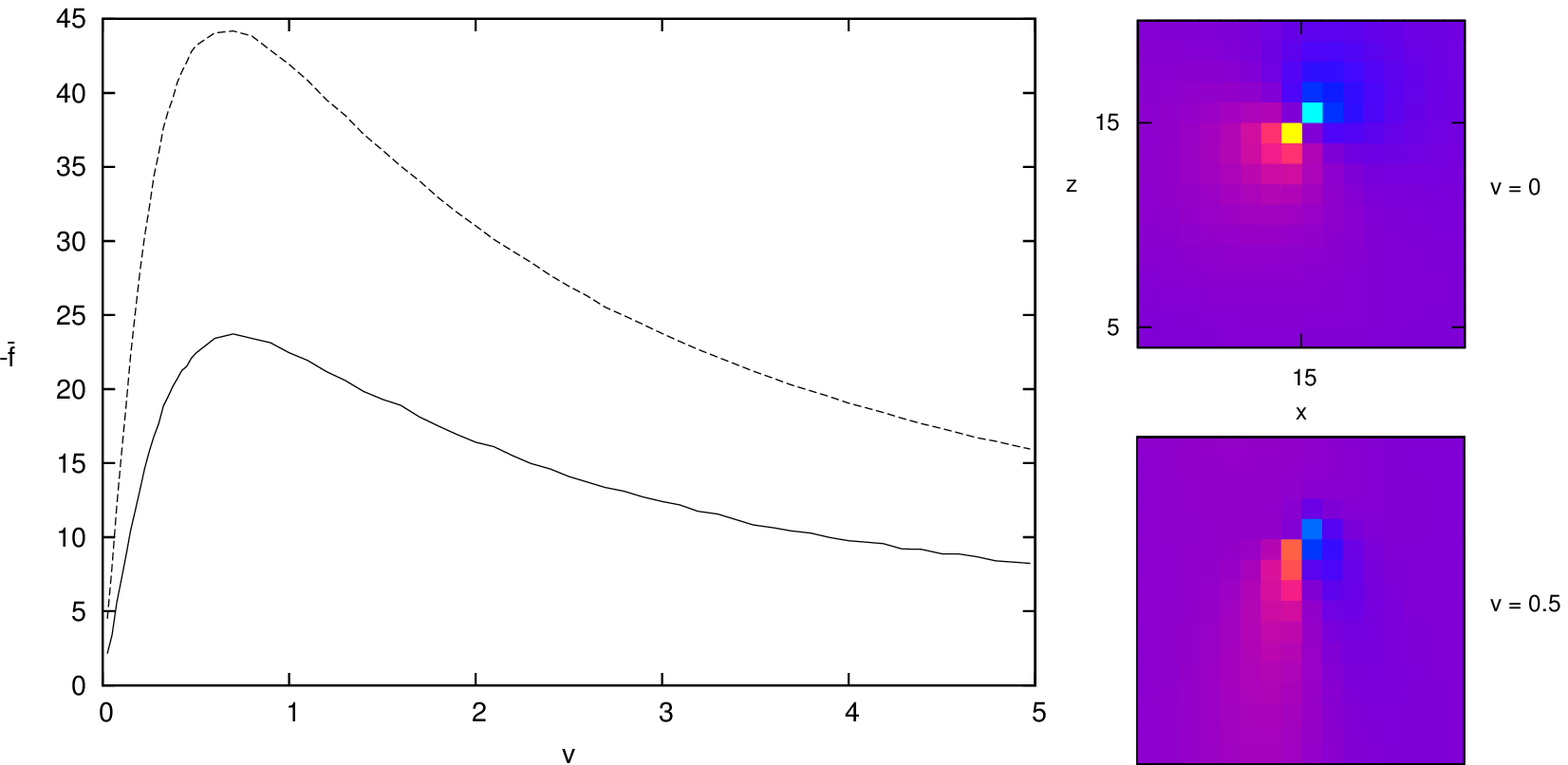}}
\end{center}
\caption{Forces for a dipole orientated at $45^o$ to the motion with Glauber dynamics. Solid line: average force perpendicular to the motion; dashed line: force parallel to the motion ($\beta=1$, $J=0.4$, $h=-6.66$). Contour plot (color online) for two values of $v$.}
\label{dip45}
\end{figure*}

\section{The one dimensional Ising model with Glauber dynamics}

In the  Section 2. we have analyzed the drag on inclusions in free fields. Our simulations in Section 3. were however for the Ising ferromagnet. We showed that the force measured in these simulations could be remarkably well fitted by a free field theory. Here we show that the drag for a point like magnetic field in the one dimensional  Ising model with Glauber dynamics \cite{gla1963} can be exactly solved within the linear response regime, where $\beta h\ll 1$, and that the force so obtained is exactly of the form predicted from the model A dynamics for the Gaussian ferromagnet. 
  
As in our simulations we will compute the symmetrized instantaneous force given by
\begin{equation}
\langle f(t)\rangle = {h\over 2}\left[ \langle S_{i_0(t)+1}\rangle - \langle S_{i_0(t)-1}\rangle \right],\label{deff}
\end{equation}
where $i_0(t) = {\rm int}(vt)$ and we have set the lattice spacing $a_0=1$. The time dependent magnetic field in this
problem can be written as 
\begin{equation}
h_k(t) = h_0\delta_{k,{i_0}(t)}.
\end{equation}
We will work in the regime where the applied
field is small and apply linear response theory to write 
\begin{equation}
\langle S_j(t)\rangle = \langle S_j(t)\rangle_0 + h\sum_{k}\int_{-\infty}^t ds \ \left\langle {\delta S_j(t)
\over \delta h_k(s)}\right\rangle_0 \delta_{k,{i_0(s)}}, \label{linrep}
\end{equation} 
where $\langle \cdot \rangle_0$ indicates averaging in the absence of the field $h$. We also 
assume that the dynamics of the system in absence of the field $h$ evolves from a statistically
homogeneous initial state such that
\begin{equation}
\langle S_i(t)\rangle_0 =\langle S_j(t)\rangle_0 , 
\end{equation}
for all $i$ and $j$. Along with Eq. (\ref{linrep}) in Eq. (\ref{deff}) this yields

\begin{equation}
\begin{array}{l}
\langle f(t)\rangle  \\ \hspace{0mm} \displaystyle{=  {h^2\over 2}\sum_{k}\int_{-\infty}^t ds \left[ \left\langle {\delta S_{i_0(t)+1}
\over \delta h_k(s)}\right\rangle_0-\left\langle {\delta S_{i_0(t)-1}
\over \delta h_k(s)}\right\rangle_0\right] \delta_{k,{i_0(t)}}}  \\ \hspace{0mm}
\displaystyle{= {h^2\over 2}\int_{-\infty}^t ds \left[ \left\langle {\delta S_{i_0(t)+1}
\over \delta h_{i_0(s)}}\right\rangle_0-\left\langle {\delta S_{i_0(t)-1})
\over \delta h_{i_0(s)}}\right\rangle_0\right] .}
\end{array}
\end{equation}

The response function for the unperturbed system is defined by
\begin{equation}
{\mathcal R}(i,j,t,s) = \left\langle {\delta S_i(t)\over\delta h_j(s)}\right\rangle_0,
\end{equation}
and for a system in thermal equilibrium we may write
\begin{equation}
{\mathcal R}(i,j,t,s) = {\mathcal R}(i-j, t-s),
\end{equation}
as we have spatial and time translation invariance and thus
\begin{equation}
\begin{array}{ll}
\langle  f (t)\rangle =  {h^2\over 2}\int_{-\infty}^t ds &[ {\mathcal R}(i_0(t)-i_0(s) +1, t-s)\\& \hspace{2mm}-
{\mathcal R}(i_0(t)-i_0(s) -1, t-s) ].
\label{int1}
\end{array}
\end{equation}
In addition, for a system
in equilibrium, one has the fluctuation dissipation theorem \cite{lip2005}
\begin{equation}
{\mathcal R}(i,j,t,s) = \beta\theta(t-s){\partial C(i,j,t,s)\over \partial s}= \beta {\partial C(i-j,t-s)\over \partial s},
\end{equation}
for $t>s$, and where
\begin{equation}
C(i,j,t,s) = \langle S_i(t)S_j(s)\rangle_0
\end{equation}
is the spin-spin correlation function. Therefore  thermal equilibrium we
find 
\begin{equation}
\begin{array}{ll}
\displaystyle{\langle f(t)\rangle= -{\beta h^2\over 2 }\int_{-\infty}^t ds{\partial\over \partial \tau}} & (C(i_0(t)-i_0(s) +1, \tau) \\ & - C(i_0(t)-i_0(s) -1, \tau))_{\tau=t-s} .
\end{array}\label{ffdt}
\end{equation}
The correlation function obeys the equation
\begin{equation}
{\partial \over \partial t}C(i,t) = -C(i,t) + {\gamma\over 2}\left(C(i+1,t)+ C(i-1,t)\right)\label{eqcorr}.
\end{equation}
where $\gamma = \tanh(2\beta J)$ \cite{gla1963}.  We now consider a system with $2L+1$ spins 
at sites $-L,\dots,\ 0\cdots,L$ and periodic boundary conditions. We define the discrete Fourier 
transform of $C$ via
\begin{equation}
C(j)= \sum_{k=-L}^{L} \tilde C(k) \exp\left({2\pi ijk\over 2L+1}\right),
\end{equation}
and thus the Fourier coefficients are given by
\begin{equation}
\tilde C(k) =  {1\over 2L+1}\sum_{j=-L}^L C(j) \exp\left(-{2\pi ijk\over 2L+1}\right).\label{ftf}
\end{equation}
The initial condition for Eq. (\ref{eqcorr}) is given by the equilibrium correlation function
\begin{equation}
C(i) = \eta^{|i|}\label{ic},
\end{equation} 
where $\eta =\tanh(\beta J)$. We can now use Eq. (\ref{eqcorr}) to express  Eq. (\ref{ffdt}) in terms 
of its Fourier representation to find
\begin{equation}
\begin{array}{l}
\displaystyle{\langle f(t)\rangle=-{\beta i h^2}\int_{-\infty}^t ds \sum_k \tilde C(k,(t-s))}\\ \hspace{10mm} \displaystyle{\times \left[1-\gamma\cos\left({2\pi  k\over 2L+1}\right)\right]} \\ \hspace{10mm} \displaystyle{\times\sin\left({2\pi  k\over 2L+1}\right)\exp\left({2\pi ik(i_0(t)-i_0(s))\over 2L+1}\right)}.
\end{array}
\end{equation}
Now in the continuum limit where $\xi \gg 1$ we write simply that  $i_0(t) = vt$ and use the solution
\begin{equation}
\tilde C(k,t) = \tilde C(k,0)\exp\left(-t\left[ 1-\gamma\cos\left({2\pi  k\over 2L+1}\right)\right]\right),
\end{equation}
to obtain
\begin{equation}
\langle f(t)\rangle= {\beta i h^2} \sum_k {\tilde C(k,0)\left[1-\gamma\cos\left({2\pi  k\over 2L+1}\right)\right] \sin\left({2\pi  k\over 2L+1}\right)
\over 1-\gamma\cos\left({2\pi  k\over 2L+1}\right)-{2\pi ik v\over 2L+1}}.
\end{equation}
The initial condtion Eq. (\ref{ic}) along with (\ref{ftf}) then gives
\begin{eqnarray}
\tilde C(k,0) &=& {1\over 2L+1} \sum_{j=-L}^{L}\eta^{|j|}\exp\left(-{2\pi ijk\over 2L+1}\right)\nonumber \\
&=& {1\over 2L+1} {1-\eta^2\over 1+\eta^2 -2\eta\cos\left({2\pi k\over 2L+1}\right)}\nonumber \\
&=& {1\over 2L+1}{1\over \cosh(2\beta J)}{1\over 1-\gamma \cos\left({2\pi  k\over 2L+1}\right)},
\end{eqnarray}
where we have taken the limit of large $L$ and assumed that $\eta <1$ (i.e. non zero temperature). Putting this all together then yields
\begin{equation}
\langle f(t)\rangle= {\beta i h^2\over \cosh(2\beta J)(2L+1)} \sum_k {\sin\left({2\pi  k\over 2L+1}\right)
\over 1-\gamma\cos\left({2\pi  k\over 2L+1}\right)-{2\pi ik v\over 2L+1}}.
\end{equation}
Now the sum can be written as an integral for $L$ large to give
\begin{equation}
\langle f(t)\rangle= {\beta i h^2\over 2\pi\cosh(2\beta J)} \int_{-\pi}^{\pi}dk\  {\sin\left( k\right)
\over 1-\gamma\cos\left(k \right)-ik v}.\label{isd}
\end{equation}
We can recover the link with the continuum models studied here if we consider the limit where 
the inverse correlation length $m\ll1$. Here we have
\begin{equation}
m = -\ln(\eta) \sim 2\exp(-2\beta J).
\end{equation}
If we take $m$ small the integral in Eq. (\ref{isd}) is dominated by its behavior at small $k$, in
addition we have $\gamma \sim 1-2\exp(-4\beta J) = 1-m^2/2$ which yields
\begin{equation}
\langle f(t)\rangle ={\beta i mh^2\over \pi} \int_{-\pi}^{\pi}dk\  {k
\over k^2 + m^2 -2ik v},
\end{equation}
which has the same form as that for model A dynamics for a Gaussian ferromagnet.

The analytical result (\ref{isd}) has been compared with a simulation; the results given Fig. \ref{isingcalc}\  show a good agreement between them. We interpret the fact that the analytical result overestimates the simulations result as a trace of a non linear response in the field $h$.

\begin{figure}
\begin{center}
\resizebox{1\hsize}{!}{\includegraphics{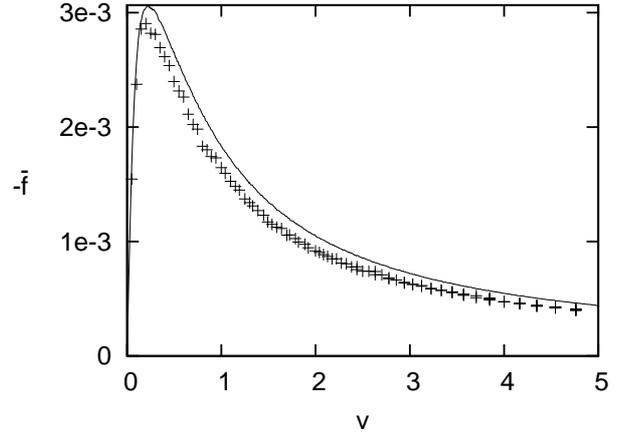}}
\end{center}
\caption{Drag force in one dimension for small field in an Ising model with Glauber dynamics (crosses) versus analytical result (\ref{isd}) (solid line); $\beta=1$, $J=1.5$, $h=0.2$.}
\label{isingcalc}
\end{figure}

\section{Application to proteins in lipid membrane}

\subsection{General analysis}

Here we will try and investigate some possible sources of drag in lipid membrane models. A way of computing the diffusion constant of an insertion, such as a protein in a lipid membrane, is via the 
Stokes Einstein relation
\begin{equation}
D= {k_B T\over \lambda_t}
\end{equation}
where $\lambda_t$ is the total friction on the protein. There are a number of possible sources of
drag in lipid membranes. The first treatment of this problem was by Saffmann and Delbr\"uck (SD) \cite{saff1975}
who computed the hydrodynamic drag by treating the low Reynolds number Navier Stokes equations for a slab of flat 2d fluid containing a solid cylindrical insertion. The movement
of the fluid sets up a hydrodynamic flow and the resulting friction on the cylinder is computed using
the stress tensor. The coupling to the bulk external fluid is very important and is essential to find a 
finite result for the drag, as a purely two dimensional treatment gives a divergent result coming from the
long range nature of hydrodynamic interactions in two dimensions.  The hydrodynamic 
drag computed by  SD  is given by
\begin{equation}
\lambda_\textrm{hydro} ={ 4\pi \eta_m\over \left[\ln\left( {\eta_m h\over \eta_w  a}\right)-\gamma\right]},
\end{equation}
where $a$ is the cylinder radius and  $\eta_m$ and $\eta_w$ are the the viscosities of the 
membrane and  surrounding fluids respectively. The term $h$ represents the height of the 
cylinder  or membrane and $\gamma \approx 0.5772$ is Euler's constant. This formula is valid in the regime where $a \gg h$, i.e. for proteins which are large relative to the membrane thickness and when
$\eta_m\gg\eta_w$.  The coupling of the 2d flow to the 3d fluid is very important in this hydrodynamic treatment, for example if there is a hard wall in the proximity of the fluid membrane, the behavior
of the diffusion constant changes to $D\sim 1/a^2$ \cite{wall}.

Recently in \cite{gam2006} a detailed experimental study, and comparison of other results in the literature, of protein diffusion constants  seems to suggest that for membrane proteins and peptides the diffusion constant scales as $D\sim1/a$ (which is consequently a much stronger dependence on the protein radius than $D\sim \ln(1/a)$ predicted by SD). In \cite{gam2006} it is suggested that the apparent 
failure of the SD formula may be due to the fact that the membrane is quite heterogeneous on small length scales and that the model of a perfect incompressible fluid is perhaps not well adapted for small
inclusions. It is also pointed out that on larger length scales thermal fluctuations and undulations may dissipate velocity gradients. Indeed extensive numerical simulations have shown that the coupling
of the protein position to local membrane curvature (and hence height fluctuations) reduces the 
diffusion constant of inclusions \cite{naj2009} and scaling arguments show  that proteins 
whose hydrophobic cores are mismatched with the equilibrium thickness of the lipid bilayer also experience additional drag forces \cite{naji2007}. The effect of mismatch is clearly seen in some of the 
experimental studies reported in \cite{gam2006}. 

In the spirit of the comments of \cite{gam2006} and the study of \cite{naji2007},  we will tentatively examine various scenarios leading to drag forces on membrane inclusions which are linearly 
coupled to physical fields in the membrane. We should bear in mind the limitations of this approach. First
it is clear that we are ignoring possible hydrodynamic flows created by the movement of the inclusion. If substantial hydrodynamics flows are established by protein movement then the order parameter field
$\phi$ (depending on its precise nature) can be expected to be convected with the flow. Clearly 
this effect is ignored in our study, however for small inclusions where the inclusion pushes past the local lipids, rather than entraining  a hydrodynamic flow, we expect this approximation to be good, at least
concerning orders of magnitude estimates. Secondly there is also a nonlinear coupling between 
inclusion position and the fluctuating field (due to the absence of the field in the region of the 
inclusion). However if the linear coupling is sufficiently strong the contribution of the mean field like term we compute here should dominate the drag due to nonlinear terms.

In  $d=2$, the friction coefficient (\ref{dragcoefiso}) reads, 
\begin{equation}
\lambda = \frac{h^2}{2\pi}\int_0^{\pi/a_c}dk\ \frac{k^3\tilde K^2(k)}{\tilde R(k)\tilde\Delta^2(k)}.
\end{equation}
where $a_c$ is the short distance cut-off scale corresponding to the larger of the two scales $a$, the insertion size, and $a_0$, the underlying
cut off for the field fluctuations. In our models, the operators will be of the forms of  Eqs. (\ref{formDelta}-\ref{formK}); in order to be more precise, we write on dimensional grounds
\begin{alignat}{2}
& \tilde \Delta(k) & = &\ \mu_\Delta a_0^{\delta'} k^{\delta'}(k^2+m^2)\ ,\\
& \tilde R(k)\tilde \Delta(k) &\ = &\ \tau_0^{-1}a_0^{\rho+\delta'+2} k^{\rho+\delta'}(k^2+m^2)\ ,  \\
& \tilde K(k) & =  &\ \mu_K a_0^\alpha k^\alpha\ ,
\end{alignat}
where $\mu_\Delta$ and $\mu_K$ are energies and $\tau_0$ is a microscopic time scale associated with the dynamics. Here $\delta'$ is simply the exponent $\delta$ of the previous sections 
when $m\neq 0$.

\par Using these expressions in the integral above, and extracting the mass dependence by setting $k=mq$, we obtain for the friction coefficient
\begin{equation}
\lambda={\tau_0\mu_K^2\over 2\pi\mu_\Delta a_0^2} h^2 (ma_0)^{2\alpha-\rho-2\delta'}g_\lambda\left(\frac{ma_c}{\pi}\right)\ ,
\end{equation}
where the function $g_\lambda$ is defined by 
\begin{equation}
g_\lambda(x)=\int_0^{1\over x} dq\ {q^{3+2\alpha-\rho-2\delta'}\over (q^2+1)^2}.
\end{equation}
In the following, we assume that this integral is not infrared divergent, i.e. that $2\delta'+\rho-2\alpha<4$; this reads $d=2>d_c$ with the critical dimension (\ref{critdim}).

It now remains to determine how the amplitude of the interaction $h$ should be computed. It is clear that
the value of $h$ should depend on the value of the size of the inclusion. A simple, semi-macroscopic,  way of doing this, proposed in \cite{dem2010}, is the following. The energy of interaction between the field $\phi$ and 
the inclusion is easily computed from the free field theory and is given by
\begin{equation}
\epsilon = -{h^2 \over 2(2\pi)}\int_0^{\pi\over a_c} dk\ {k\tilde K^2(k)\over\tilde \Delta(k)}.\label{eqe}
\end{equation}
We now expect that $\epsilon$ is a function of $a$ and that for small $a$
\begin{equation}
\epsilon(a) \sim -2\pi \gamma_I a - \pi \sigma_I a^2 \label{line}
\end{equation}
where $\gamma_I$ and $\sigma_I$ are effective (negative) line and surface tensions for the inclusion in the membrane due to the interaction with the field $\phi$. Now if we assume that $a$ is small we can neglect the surface tension term and equating Eq. (\ref{eqe}) and Eq. (\ref{line}) we find 
\begin{equation}
2\pi\gamma_I a = {h^2\over 2 (2\pi)}{\mu_K^2(a_0 m)^{2\alpha-\delta'}\over \mu_\Delta}g_\epsilon\left({ma_c\over \pi}\right)
\end{equation}
where
\begin{equation}
g_\epsilon(x) = \int_0^{1\over x}  dq\ {q^{2\alpha-\delta'+1}\over q^2 +1}.
\end{equation}
Using the resulting expression for $h$ in terms of $\gamma_I$ and $a$ we then obtain
\begin{equation}
\lambda = {4\pi\gamma_I a\tau_0 g_{\lambda}({ma_c\over \pi})\over a^2_0 (ma_0)^{\rho+\delta'} g_\epsilon({ma_c\over \pi})}.
\end{equation}

Before examining a number of models of proteins inserted into membranes we will explore a few general consequences of the above expression. Given that we expect $a$  and $a_0$ to be small
we should consider the functions $g_\lambda$ and $g_\epsilon$ in the limit $x\to 0$. The integrals defining these functions are finite (and thus independent) of $a$ in the cases where
$2\alpha-\rho-2\delta' <0$ and $2\alpha-\delta' <0$ respectively. In this case we find the 
generic behavior
\begin{equation}
\lambda \approx {4\pi\gamma_I a\tau_0 g_{\lambda}(0)\over a^2_0 (ma_0)^{\rho+\delta'} g_\epsilon(0)},
\end{equation}
In this scenario we see that the dependence of the friction coefficient on the inclusion size is
always linear and it has a strong dependence on the correlation length of the field $\phi$. The scaling of the friction coefficient with the inclusion size is $\lambda \sim a$, if this drag
dominates all other sources of drag  application of the Stokes-Einstein  relation 
gives
\begin{equation}
D\sim {1\over a}.
\end{equation}
Note that we will also recover this dependence on $a$ if it is the case that $a<a_0$, i.e. the underlying
cut-off of the field $\phi$ is greater than the one corresponding to the inclusion size.  
Note that apart from the hydrodynamic case  where $\rho=-1$, $\rho$ is positive or zero, therefore
if  $2\alpha-\delta' <0$, then  $2\alpha-\rho-2\delta' <0$ also. In the case where
$2\alpha-\rho-2\delta' >0$ and which in most cases will also imply that  $2\alpha-\delta' >0$, the integrals defining both $g_\lambda(x)$ and $g_\epsilon(x)$ will diverge and we find
\begin{eqnarray}
g_\lambda(x) &\sim& {1\over 2\alpha-\rho-2\delta'}{1\over x^{2\alpha-\rho-2\delta'} }\\
g_\epsilon(x) &\sim& {1\over 2\alpha-\delta'}{1\over x^{2\alpha-\delta'} }
\end{eqnarray}
and thus
\begin{equation}
\lambda \approx  {4\pi\gamma_I a\tau_0 \over a^2_0} {2\alpha-\delta'\over 2\alpha-\rho
-2\delta'}\left({a\over \pi  a_0}\right)^{\rho+\delta'}.
\end{equation}
Again if this drag dominates, it gives a diffusion coefficient scaling with inclusion size as
\begin{equation}
D\sim {1\over a^{\rho+\delta'+1}}.
\end{equation}
In this case we see a different dependence on the inclusion size $a$, the physics of the problem is controlled by short distance behavior and the drag is independent of the correlation length $\xi=1/m$
of the fluctuating field. A final possible case is where $2\alpha-\rho-2\delta' <0$ but $2\alpha-\delta' >0$, in which case we find 
\begin{equation}
\lambda \approx  \frac{4\pi\gamma_I a\tau_0 g_\lambda(0)(ma)^{2\alpha-\delta'}}{a^2_0(ma_0)^{\rho+\delta'}} {2\alpha-\delta'\over \pi^{2\alpha-\delta'} },
\end{equation}
this is a particularly interesting case as the friction coefficient has a strong dependence on both the 
correlation length of the field $\phi$ and on the inclusion size.  Again if this drag dominates the diffusion
constant scaling with inclusion size is 
\begin{equation}
D\sim {1\over a^{2\alpha+1-\delta'}}
\end{equation}

We now discuss some specific models.

\subsection{Insertion with curvature coupled to membrane height fluctuations}

In \cite{naj2009} the  authors  numerically investigated the diffusion constant of inclusions in model
membrane systems where the inclusion tends to impose a preferred local curvature on the 
membrane. In their model a quadratic coupling was also considered. Here we consider 
the model with a simple linear coupling. The inclusion is coupled to the membrane height fluctuations, 
the Gaussian Hamiltonian of which $\Delta$ is given by Eq. (\ref{helfrich}), 
the dynamical operator $R$ is given by Eq. (\ref{hdf})  and $K$ by Eq. (\ref{kcurv}). After Fourier transforming, we obtain $\tilde\Delta(k)=\kappa k^2(k^2+m^2)$ (with $m=\sqrt{\sigma/\kappa}$), $\tilde R(k)=(4\eta k)^{-1}$ and $\tilde K(k)=-k^2$.
In our general notation we  have $\delta'=2$, $\alpha=2$ and $\rho=-1$ and we are in the case where
$2\alpha-\delta' = 2$ and $2\alpha-\rho-2\delta' =1$. In terms of the physical parameters of this model
the drag coefficient for small insertion size $a$ is thus given by
\begin{equation}
\lambda = {32\eta \gamma_I a^2\over \kappa},
\end{equation}
and the dominance of this drag would imply that
\begin{equation}
D\sim {1\over a^2}.
\end{equation}
We thus see that this result is quite insensitive to the correlation length of the height fluctuations
(and thus the surface tension) assuming that they are large with respect to the insertion size.

\subsection{Insertion coupled to a non-conserved order parameter}

An insertion such as a protein or peptide can couple to various physical fields in a lipid membrane other than the height fluctuations. For instance in a lipid monolayer local tilt angles of the lipid heads or tails
may be changed by the presence of an inclusion. Also if there are several lipid phases such as liquid gel and solid,  the inclusion may prefer to be in one of these phases. This general idea can be modelled
by assuming that the inclusion couples linearly to the order parameter representing one of these fields. 
The simplest Hamiltonian for this order parameter has the form of that for the Gaussian ferromagnet where  $\tilde\Delta(k)\propto k^2+m^2$. The simplest diffusive dynamics is given by model A dynamics with $\tilde R(k)\propto 1$ and a linear coupling to the field $\phi$  gives $\tilde K(k)=1$. This  dynamical model does not conserve the integrated field as there is no reason that it should be conserved.  
Here, in our general notation, we  have $\delta'=0$, $\alpha=0$ and $\rho=0$ and we are thus in the case where $2\alpha-\delta' = 0$ and $2\alpha-\rho-2\delta' =0$. We see that we are in the marginal case for both functions $g_\lambda$ and $g_\epsilon$. Furthermore we can identify the  time scale
$\tau_0$ using the diffusion constant for the dynamics of the field $\phi$, $D_0$,  via $D_0\tau_0 = a_0^2$, where $a_0$ is the lipid size and $D_0$ can be estimated from the lipid translational diffusion constant or lipid rotational diffusion constant, depending on  field in question
(for instance if the field in question describes the  orientational order of the lipids, then the 
lipid rotational diffusion constant could be used to give the appropriate time scale). Here for small $x$ we find 
\begin{equation}
g_\lambda(x) \approx -\ln(x) \ \ {\rm and }\ \  g_\epsilon(x) \approx -\ln(x)
\end{equation}
which leads to 
\begin{equation}
\lambda \approx {4\pi\gamma_I a \over D_0 },
\end{equation}
and gives an estimation of the insertion diffusion constant:
\begin{equation}
D\approx {k_BT D_0\over 4\pi \gamma_I  a}.
\end{equation}

\subsection{Insertion coupled to a conserved order parameter}

The insertion may also be coupled to a conserved field, describing, for example, the local lipid composition in the case where there are several lipid types. We take the same $\Delta$ and $K$ operator to describe the energy of the order parameter describing local chemical composition. 
However we now use a conserved dynamics: $R$ is set by (\ref{mbd}), giving $\tilde R(k) 
\propto k^2$.  We thus have $\delta'=0$, $\alpha=0$ and $\rho=2$, which gives
$2\alpha-\delta' = 0$ and $2\alpha-\rho-2\delta' =-2$. The function $g_\epsilon$ is unchanged
but we find $g_\lambda \approx 1/2$ which gives
\begin{equation}
\lambda = {2\pi\gamma_I a\over D_0a^2_0 m^2 \ln({\pi\over ma})}
\end{equation}
where we have again used $D_0\tau_0 = a_0^2$, and where $D_0$ can be estimated from the lipid
translational diffusion constant.  This leads to the estimate
\begin{equation}
D= {k_B TD_0a^2_0 m^2 \ln({\pi\over ma})\over 2\pi\gamma_Ia}
\end{equation}
for the protein diffusion constant. We should note that even though there is a logarithmic correction,
we would expect to experimentally measure $D\sim 1/a$ as the logarithmic term would require
decades of length scales (thus leaving the realm of validity of the calculation) to detect. Interestingly here, in contrast with the previous models, we should see a strong dependence of $D$ on the 
correlation length of the field.

\section{Conclusions}
We have seen that inclusions which are linearly coupled to classical fields with dissipative 
dynamics are subject to drag forces which exhibit a rather rich behavior. Notably the drag force
is a non monotonic function of the inclusion velocity $v$. Generically the force is a linear function
of the velocity at small $v$ and is characterized by a friction coefficient $\lambda$. The force then
attains a maximum value, and for large $v$ decays as $1/v$. The force is physically generated by
the deformation of the polarization profile of the field about the inclusion by the inclusions motion.
This phenomena is analogous to the way in which electron dynamics is renormalized by their
associated polaron in solid state physics \cite{polaron}. The linear coefficient of friction $\lambda$ is
of particular importance as it can be used to estimate diffusion constants via the Stokes-Einstein relation and because it can exhibit divergent behavior when the corresponding field theory is critical, i.e.
has a diverging correlation length, for example at a critical demixing transition.

The results we have presented are valid for free or Gaussian field theories. We were able to compute the drag for the one dimensional Ising model for a weak inclusion-field 
interaction but it would be interesting to go beyond the Gaussian approximation to understand
the physics of drag forces for general interacting field theories. Having said this we note that
the Gaussian model does seem to capture most of the phenomenology seen in our simulations
of the one and two dimensional Ising model.

Finally the experimental measurement  of these drag forces presents an interesting challenge.
It may be possible to carry out experiments using atomic force microscopy or magnetic force 
microscopy if the interaction between the microscopic tip and the surface can be sufficiently
well characterized. Also, the thermal or critical Casimir force predicted by Fisher and
de Gennes \cite{fi1978} has recently been successfully measured \cite{he2008} in a binary 
fluid mixture at criticality. It may be that the technical advances made to carry out this measurement, the chemical treatment to tune the interaction between the fluid components and surfaces, and the optical
force measurements could be applied to the study of the drag problem. We note also that, beyond 
measurements of the average force, it would also be interesting if the predictions made here about
force fluctuations could be verified experimentally.

\noindent {\bf Acknowledgements:} DSD thanks the Institut Universitaire de France for financial support. We would also like to thank S.  Ramaswamy for helpful comments and pointing out a number of useful references. 

%


\begin{thebibliography}{}
\bibitem{wein} S. Weinberg, {\em The Quantum Theory of Fields}, Volume 1 (Cambridge University Press, Cambridge) (2005).
\bibitem{gou1993}M. Goulian, R. Bruinsma, and P. Pincus, Europhys. Lett. {\bf 22}, 145 (1993).
\bibitem{sac1995} E. Sackmann in {\em Structure and Dynamics of Membranes, From Cells to Vesicles}
Eds. R. Lipowksy and E. Sackmann (Elsevier Science BV, Amsterdam) (1995).
\bibitem{nem}{H. Imura and K. Okano, Phys. Lett. {\bf 42}A, 403 (1973) ; G. Ryskin and M. Kremenetsky, Phys. Rev. Lett. {\bf 67}, 1574Ð1577 (1991).}
\bibitem{dub1983}E. Dubois-Violette, E. Guazzelli and J. Prost, Philos. Mag. A {\bf 48}, 727 (1983).
\bibitem{lub1986}T. C. Lubensky, S. Ramaswamy and J. Toner, Phys. Rev. B 33, 7715 (1986).
\bibitem{fus2008}C. Fusco, D.E. Wolf and U. Nowak, Phys. Rev. B, {\bf 77}, 174426 (2008); M.P. Magiera, L. Brendel, D.E. Wolf and U. Nowak, Eur. Phys. Lett. {\bf 87}, 26002 (2009).
\bibitem{dem2010}{V. D\'emery and D.S. Dean, Phys. Rev. Lett. {\bf 104}, 080601, (2010).}
\bibitem{saff1975}P.G. Saffmann and M. Delbr\"uck, Proc. Natl. Acad. Sci USA {\bf 72}, 3111 (1975).
\bibitem{dego2009}{D.S. Dean and A. Gopinathan, J. Stat Mech L08001 (2009).}
\bibitem{hel1973}{W. Helfrich, Z. Naturforsch. {\bf 28}c, 693 (1973).}
\bibitem{chai2000}{P.M. Chaikin and T.C. Lubensky, Principles of Condensed Matter Physics (
Cambridge University Press, Cambridge) (2000).}
\bibitem{comm}{Solving the problem in Fourier space ensures that derivatives so computed
are averaged over their left and right values (Dirichlet's theorem), this thus corresponds to our 
method of computing the force via the difference in energy of a move forwards and a move backwards.}
\bibitem{vor}{L. P. Gor'kov and N. B. Kopnin, Sov. Phys. Usp. {\bf 18}, 496 (1975);  A.T. Dorsey, Phys. Rev. B {\bf }46, 8376 (1992); R. A. Simha and S. Ramaswamy, Phys. Rev. Lett. {\bf 83}, 3285Ð3288 (1999).}
\bibitem{polaron}L. D. Landau, Phys. Z. Sowjetunion, {\bf 3}, 644 (1933); H. Fr\"ohlich, Adv. in Phys. {\bf 3}, 325 (1954).
\bibitem{gla1963}{R.J. Glauber, J. Math. Phys. {\bf 4}, 294 (1963).}
\bibitem{lip2005}{E. Lippiello, F. Corberi and M. Zannetti, Phys. Rev. E {\bf 71}, 036104 (2005).}
\bibitem{wall}{E. Evans and E. Sackmann, J. Fluid Mech. {\bf 194}, 553 (1988); R. Merkel, E. Sackmann and E. Evans, J. Phys.  France {\bf 50}, 1535 (1989).}
\bibitem{gam2006} Y. Gambin et. al., Proc. Natl. Acad. Sci. USA, 103, 2089 (2006).
\bibitem{naj2009} A. Naji, P.J. Atzberger and F.L.H. Brown, Phys. Rev. Lett. {\bf 102}, 138102 (2009).
\bibitem{naji2007}A. Naji, A.J. Levine and P.A. Pincus, Biophys. J. {\bf 93}, L49 (2007).

\bibitem{fi1978}{M.E. Fisher and  P.-G. de Gennes, C. R. Acad. Sci. Paris B {\bf 287}, 207 (1978).}
\bibitem{he2008}{C. Hertlein, L. Helden, A. Gambassi, S. Dietrich and  C. Bechinger, Nature {\bf 451}, 172 (2008).}


\end{thebibliography}
\end{document}